\documentclass[prl,twocolumn,groupedaddress]{revtex4}
\newcommand{\Rv}{{\bf R}}
\newcommand{\rv}{{\bf r}}
\newcommand{\xv}{{\bf x}}

\newcommand{\qv}{{\bf q}}
\newcommand{\nv}{{\bf n}}
\newcommand{\kv}{{\bf k}}

\newcommand{\ppi}{{\partial_i}}
\newcommand{\ppj}{{\partial_j}}
\newcommand{\pz}{{\partial_z}}

\usepackage[dvips]{graphicx}

\begin{document}
\title[]{Universal Elasticity and Fluctuations of Nematic Gels}
%Spontaneously
%Orientationally-Ordered Solids}
\author{Xiangjun Xing}
\author{Leo Radzihovsky}
%\email[]{Xing@ucsub.colorado.edu}
%\thanks{}
\affiliation{Department of Physics, University of Colorado,
   Boulder, CO 80309}


\date{\today}

\begin{abstract}
  
  We study elasticity of spontaneously orientationally-ordered
  amorphous solids, characterized by a vanishing transverse shear
  modulus, as realized for example by nematic elastomers and gels.  We
  show that local heterogeneities and elastic nonlinearities conspire
  to lead to anomalous nonlocal universal elasticity controlled by a
  nontrivial infrared fixed point.  Namely, at long scales, such solids
  are characterized by universal shear and bending moduli that,
  respectively, vanish and diverge at long scales, are universally
  incompressible and exhibit a universal negative Poisson ratio and a
  non-Hookean elasticity down to arbitrarily low strains. Based on
  expansion about five dimensions, we argue that the nematic order is
  stable to thermal fluctuation and local heterogeneities down to
  $d_{lc}<3$.

\end{abstract}

%\maketitle must follow title, authors, abstract and \pacs
\maketitle

%\section{}
%\label{}
%\subsection{}
%\subsubsection{}

%\narrowtext
%\widetext

%\section{Introduction}
%\section{Model}
%An important question is how many independent parameters 
%we need to describe the elastic energy cost for uniform 
%deformation of nematic elastomer? The answer is 9?

Liquid crystal elastomers and gels --- weakly crosslinked networks of
liquid crystal polymers --- combine the electro-optic response and
thermodynamic phase behavior of liquid crystals with the mechanical
advantages of solids, such as rubber and plastics. They therefore hold
considerable technological potential. These materials exhibit rich
interplay between orientational order and network elasticity that
leads to many unusual properties not found in conventional liquid
crystals or in conventional rubber\cite{review_elastomers}.  The most
striking of these is the vanishing stress, $\sigma_{ij}$ in response
to a finite strain, $u_{ij}$ applied transversely to the
spontaneous\cite{semisoft} uniaxial distortion that develops below the
isotropic-nematic (IN) transition. Much progress had been made in
understanding these materials both from the neo-classical theory of
rubber\cite{review_elastomers} and from more general symmetry-based
elastic formulation\cite{GL,LMRX}. Many of the intriguing properties
of nematic elastomers can be traced back to the existence of novel
nemato-elastic Goldstone mode\cite{softmodes,Olmsted}, associated with
a spontaneous breaking of rotational symmetry\cite{GL,LMRX} of the
amorphous polymer matrix.  Although considerable progress had been
made\cite{review_elastomers}, most of the analyses had been limited to
mean-field treatments. We have recently developed a fully {\em
nonlinear} elastic theory of nematic elastomers\cite{XRunpublished},
that, for one, allows us to assess the effects of thermal
fluctuations\cite{elastomer_thermal}.  Furthermore, while {\em
statistically} homogeneous and isotropic\cite{semisoft}, elastomers
are {\em locally} quite heterogeneous\cite{heterogeneity}. It is
essential to study the role that network heterogeneity plays in
determining macroscopic properties of liquid crystal elastomers, and
this is the goal of the present Letter.

%Our questions of interest and approach are intimately related to the
%recent work on conventional liquid crystal confined inside a rigid
%porous matrix such as for example
%aerogel\cite{Zumer,RT,Science,MIT,Feldman}.  The main qualitative
%difference, however, is the compliance of the elastomer network, that
%unlike aerogel distorts substantially in response to the liquid
%crystal order. Therefore, in addition to the questions about the
%liquid crystal order in the presence of such heterogeneous elastic
%network, we are also concerned with the complementary question of the
%effect of liquid crystal order on the elastic properties of such
%liquid crystal rubber. 

We find that on scales longer than $\xi^z_{NL}\sim K^2/\Delta$, (with
$K$ an effective Frank nematic modulus and $\Delta$ a measure of
heterogeneity) even arbitrarily weak heterogeneity qualitatively
modifies liquid crystal and elastic properties of a nematic elastomer
relative to those of the ideal homogeneous and isotropic
one\cite{GL,LMRX}. In particular we find that macroscopically a
nematic elastomer exhibits a nonlocal elasticity, characterized by
shear moduli, that vanish as universal power-laws of system size,
implying a host of exotic elastic behavior: (1) a {\em non}-Hookean
elasticity $\sigma_{zz}\sim (u_{zz})^\delta$ (with $\delta > 1$
universal but geometry-dependent), that extends down to arbitrarily
weak strains, even along the uniaxial axis $\hat{\bf
z}$\cite{semisoft}, contrasting strongly and qualitatively with the
{\em linear} longitudinal stress-strain relation of idealized
heterogeneity-free elastomers; (2) a crossover to linear stress-strain
relation for strain smaller than a critical strain $u_{zz}^c\sim
H^{2/\delta}$ set by an external aligning (e.g., magnetic) $H$ field;
(3) a nematic Frank modulus that diverges as a power-law of the
minimum of lengths associated with system size, stress or aligning
field; (4) universal ratios of elastic moduli, with a negative Poisson
ratio in the plane transverse to the nematic axis $\hat{z}$, and (5) a
macroscopic incompressibility, independent of microscopic elastic
moduli, even for elastomers that are compressible on short scales.  In
particular, (4) and (5) predict that a strain $u_{xx}$, applied {\em
transversely} to the nematic axis\cite{comment_n0} must be accompanied
by strains $u_{yy} = 5 u_{xx}/7$ and $u_{zz} = -12 u_{xx} / 7$, and
strain $u_{zz}$, applied {\em along} the uniaxial axis will result in
$u_{xx} = u_{yy} = - u_{zz}/2$, independent of any other microscopic
details.

We also find that because of the compliance of the elastomer network,
the nematic order is far more robust to the disordering effects of the
local random torques by network's heterogeneities. In fact, a
renormalization group analysis and expansion about five dimensions
suggests that orientational order of nematic elastomers is stable to
weak network heterogeneity\cite{stableN}, in strong contrast to
nematics confined in {\em rigid}
gels\cite{Gingras,RT_smectic,Feldman}.  

Although such quenched
disorder and thermally-driven elastic anomalies, controlled by a
nontrivial low temperature fixed point have been previously predicted
in a variety of other systems\cite{membranes,RT_smectic,RT_other}, to
our knowledge, nematic elastomers are a first example of a
three-dimensional solid, where these exotic theoretical predictions
can be {\em directly} experimentally tested.

%\begin{figure}[h] 
%\begin{center}
%\includegraphics[width=7cm,height=4cm]{deformation.eps}
%\caption{Simple experiment to verify the universal ratio between
%transverse and longitudinal shear moduli for small deformation.  The
%deformation is exaggerated to make it apparent.} 
%\label{flowdiagram}
%\end{center} 
%\end{figure}

As discussed in great detail in Refs.\ \onlinecite{GL,LMRX}, most of
the properties of nematic elastomers can be captured by a purely
elastic description in terms of a Lagrangian strain tensor
$u_{\alpha\beta} = \partial_\alpha \Rv \cdot\partial_\beta \Rv -
\delta_{\alpha\beta}$, with the nematic order parameter
$Q_{\alpha\beta}=S(\hat{n}_\alpha\hat{n}_\beta -
\delta_{\alpha\beta}/3)$ ($S$ and $\hat{\bf n}$, respectively, the
magnitude and the unit director characterizing the nematic order)
integrated out.  In this formulation the IN transition is signaled by
a spontaneous uniaxial distortion characterized by a strain
$u^0_{\alpha\beta}\propto Q_{\alpha\beta}$, that takes place when the
effective shear modulus drops below a critical value.

To study the properties of such an orientationally-ordered solid, it is
convenient to express its elasticity in terms of the strain tensor (and
its gradients) measured relative to the spontaneously uniaxially
strained state.  The resulting elastic theory resembles that of a
conventional uniaxial solid, but with an essential difference, that the
shear modulus $\mu_{z\perp}$, associated with distortions transverse to
the uniaxial axis identically vanishes\cite{GL,LMRX}.  This important
feature captures, at the harmonic level, the soft elastic mode
associated with the underlying rotational invariance of the high
temperature {\em isotropic} elastomer solid from which the uniaxial
state {\em spontaneously} emerges.

Of course real elastomers are only {\em statistically} homogeneous and
isotropic. As discussed in closely related
contexts\cite{RT_smectic,membranes,RT_other}, at long length scales,
the elastomer network heterogeneities lead to quenched local random
stress $\sigma_{\alpha\beta}(\xv)$ and anisotropy ${\bf g}(\xv)$
fields\cite{no_rf}
\begin{equation}
{\mathcal H}_{R} =-u_{\alpha\beta}\sigma_{\alpha\beta}(\xv)-\left({\mathbf
g}(\xv)\cdot {\mathbf n}\right)^2,
\end{equation}
that locally distort the elastomer relative to an idealized
microscopically homogeneous and isotropic state.

As we will show explicitly below, because of the soft mode (vanishing
$\mu_{z\perp}$), such random fields (as well as thermal fluctuations)
lead to large elastomer distortions at which some of the nonlinear
elastic terms become comparable to harmonic ones. It is therefore
essential to capture the full underlying (target and reference
spaces\cite{LMRX}) rotational invariance (soft mode) of the nematic
elastomer. As we have recently demonstrated, this is quite nontrivial
and requires a proper treatment of the nonlinear parts of the
Lagrangian strain, as well as keeping of the cubic and quartic terms
in Lagrange strain tensor\cite{elastomer_thermal}. The upshot of that
analysis is that for weak heterogeneity, at long length scales elastic
and orientational properties of a nematic elastomer\cite{stableN} are
captured by an elastic Hamiltonian
\begin{eqnarray}
{\mathcal H} &=& \frac{1}{2} \left[
B_z w_{zz}^2 + \lambda w_{ii}^2 + 2 C w_{zz} w_{ii}
+ 2 \mu w_{ij} w_{ij} \right.\nonumber\\
 &+& \left. K (\nabla_{\perp}^2 u_z)^2 \right]
- {\bf \sigma}(\xv) \cdot \nabla_{\perp} u_z,
\label{Hamiltonian_disorder}
\end{eqnarray}
where 
\begin{eqnarray}
w_{zz} &=&  \pz u_z + \frac{1}{2} (\nabla_{\perp} u_z)^2,\nonumber\\
w_{ij} &=&   \frac{1}{2} (\ppi u_j + \ppj u_i ) 
- \frac{1}{2}\ppi u_z \ppj u_z,
\end{eqnarray}
are the rotationally-invariant smectic-like and columnar-like
nonlinear strain tensors relative to the uniaxial state, $i$ and $j$
are summed over indices $x$ and $y$ transverse to the uniaxial axis
$z$, and $\sigma_i(\xv)\approx\frac{1}{2}\sigma_{iz}(\xv) + 2
g_i(\xv)g_z(\xv)$, that we take to be zero-mean Gaussian quenched
random field, with spatially independent correlation function
\begin{equation}
\overline{\sigma_i(\xv) \sigma_j (\xv')} = \Delta \delta_{ij} \delta(\xv-\xv').
\end{equation}

Our goal here is to study long scale properties of $\mathcal H$. The
harmonic correlation functions $\overline{\langle
  u_\alpha(\qv)u_\beta(\qv')\rangle}_0
=\tilde{G}^0_{\alpha\beta}(\qv)(2\pi)^d\delta^d(\qv+\qv')$
of the phonon fields $u_\alpha=(u_i,u_z)$ can be easily calculated,
\begin{equation}
\tilde{G}^0_{\alpha\beta}(\qv) = G_{\alpha\beta}^0(\qv) + 
\Delta q_{\perp}^2 G_{\alpha z}^0(\qv) G_{\beta z}^0(\qv),
\label{uu_corr}
\end{equation}
where $G_{\alpha\beta}^0(\qv)$ are harmonic correlation functions of
an idealized homogeneous and isotropic elastomer, given by
\begin{eqnarray}
G^0_{zz}(\qv) &=& \frac{1}{B_z (1 - \rho)q_z^2 + K q_{\perp}^4},\\
G^0_{zi}(\qv) &=& -\frac{C q_z q_i}
 {(\lambda + 2 \mu) q_{\perp}^2 (B_z (1 - \rho)q_z^2 + K q_{\perp}^4)} ,\\
G^0_{ij}(\qv) &=& \frac{B_z q_z^2+K q_{\perp}^4}
{(\lambda + 2 \mu) (B_z (1 - \rho)q_z^2 
+ K q_{\perp}^4)} \frac{q_i q_j}{q_{\perp}^4}\nonumber\\
&+& \frac{1}{\mu q_{\perp}^2} (\delta_{ij} - \frac{q_i q_j}{q_{\perp}^2}),
\end{eqnarray}
and $\rho = C^2/B_z (\lambda+2 \mu)$ is a dimensionless ratio. In
Eq.\ref{uu_corr}, the second term describes the dominant
heterogeneity-induced (frozen) deformations, relative to the ideal
nematic elastomer state, with the first giving thermal fluctuations
about this ground state.

The validity of the perturbation theory in heterogeneity and elastic
nonlinearities is controlled by the fluctuation of the phonon fields.
At a harmonic level, at long scales, a representative real-space
distortion is given by
\begin{eqnarray}
\overline{\langle u_z(\rv)^2 \rangle}
% &\approx& \int d^d q  \frac{\Delta q_{\perp}^2}
%{\left( B_z (1 - \rho)q_z^2 + K q_{\perp}^4 \right)^2}\nonumber\\
&=& \mbox{Const.} \Delta \left( \frac{B_z(1-\rho)}{K^5}\right)
^{\frac{1}{2}} L_{\perp}^{5-d},
\end{eqnarray}
%with the quenched random contribution dominating over the thermal one.
The divergence of these distortions with the size of the system
$L_{\perp,z}$ for $d<5$ signals the breakdown of harmonic elasticity
on scales longer than nonlinear length scales $\xi^{\perp,z}_{NL}$,
which, in 3d are given by
\begin{equation}
\xi^{\perp}_{NL}\approx 
\frac{K^{5/4}}{(B_z (1-\rho))^{1/4} \Delta^{1/2}}\;,\qquad
\xi^z_{NL}\approx K^2/\Delta.
\label{xiNL}
\end{equation}
%We apply standard replica trick to model Eq.\ref{Hamiltonian_disorder} so
%as to restore the translational symmetry. The resulting ``n replicated''
%Hamiltonian is 
%\begin{eqnarray}
%{\mathcal H}[\uv^{\alpha}] &=& \sum_{\alpha} \left[
%B_z (w_{zz}^{\alpha})^2 + \lambda (w_{ii}^{\alpha})^2 
%+ 2 C w_{zz}^{\alpha} w_{ii}^{\alpha}
%+ 2 \mu w_{ij}^{\alpha} w_{ij}^{\alpha}\right.\nonumber\\
%&+& \left. K (\nabla_{\perp}^2 u_z^{\alpha})^2 \right] -
%\Delta \sum_{\alpha, \beta }\nabla_{\perp} u_z^{\alpha}
%\cdot \nabla_{\perp} u_z^{\beta}.\label{replicated_Hamiltonian}
%\end{eqnarray}

\begin{figure}[!htbp]
\begin{center}
  \includegraphics[width=7cm,height=3cm]{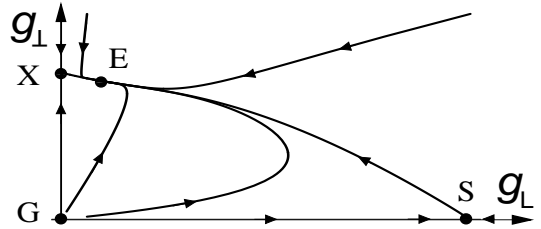}
\caption{Flow diagram for dimensionless couplings $g_L(l)$ and
$g_{\perp}(l)$.}
\label{flowdiagram}
\end{center}    
\end{figure}

To assess the physical consequence of these divergences, we employ a
momentum-shell renormalization group (RG), controlled by an expansion
in $\epsilon=5-d$. 
%(analytically continuing our model to $d$ dimensions).  
It is convenient to work with the moduli that renormalize multiplicatively,
\begin{eqnarray}
B &=& \frac{1}{4} (B_z + B_{\perp} + 2 C),\\
\mu_L &=& \frac{1}{4} (B_z + B_{\perp} - 2 C),\\
\tilde{C} &=& \frac{1}{4} (B_z - B_{\perp}),
\end{eqnarray} 
where $B$ is the overall bulk modulus, $\mu_L$ the longitudinal shear
modulus and $\tilde{C}$ the cross coupling.  The results of the RG
coarse-graining procedure are summarized by flow equations for the
effective elastomer parameters on length scale $a e^l$ ($a$ a cutoff
set by network's mesh size)
\begin{eqnarray}
\frac{d B}{d l} &=& (d+3-3 \omega - \eta_B) B ,\nonumber\\
\frac{d \tilde{C}}{d l} &=& (d+3-3 \omega- \eta_{\tilde{C}}) 
   \tilde{C} ,\nonumber\\
\frac{d \mu_L}{d l} &=&  (d+3-3 \omega- \eta_L) \mu_L,\nonumber\\
\frac{d \mu}{d l} &=& (d+3-3 \omega - \eta_{\perp}) \mu,\nonumber\\
\frac{d K}{d l} &=& (d-1-\omega+\eta_K) K, \nonumber\\
\frac{d \Delta}{d l} &=& (d+1-\omega + \eta_{\Delta})\Delta,
\end{eqnarray}
where,
\begin{eqnarray}
\eta_B&=&\frac{3}{2} \rho_2^2 g_L,\nonumber\\
\eta_{\tilde{C}}&=&\eta_L = \frac{3}{2}g_L,\nonumber\\
\eta_{\perp} &=& \frac{1}{16} g_{\perp},\nonumber\\
\eta_{\Delta}&=&\frac{8\left( 1 - \rho_2^2 \right)g_L^2 
+ 3\left( 1 + \rho_1 + 2\rho_2{\sqrt{\rho_1}}\right)g_L{g_{\perp}}}
{64\left( 1 + \rho_1 - 2\rho_2\sqrt{\rho_1}\right)
       {g_L} + 96\rho_1{g_{\perp}}},\nonumber\\
\eta_K &=& 2 \eta_{\Delta} 
+\frac{{g_{\perp}}\,\left( 4\,\left( 2 + 3\,\rho_1 
      - 5\,\rho_2\sqrt{\rho_1} \right)
         \,{g_L} + 5\,\rho_1\,{g_{\perp}} \right) }{4\,
    \left( 2\,\left( 1 + \rho_1 - 2\,\rho_2\sqrt{\rho_1}
    \right) \,{g_L} + 
      3\,\rho_1\,{g_{\perp}} \right) } .\nonumber
\end{eqnarray}
The physics is controlled by the flows of two dimensionless coupling,
$g_L(l)$ and $g_{\perp}(l)$, and two ratios $\rho_1(l)$ and
$\rho_2(l)$ defined as:
\begin{eqnarray}
g_L &=& \frac{\Delta {{\mu }_L}}{K^3} 
    {\sqrt{\frac{K\left( 2B - 4\tilde{C} + 3\mu  + 
            2{{\mu }_L} \right) }{B (3 \mu + 8 \mu_{L})+
  3 \mu (\mu_L + 2 \tilde{C}) - 8 \tilde{C}^2 }}},\qquad\label{gL}\\
g_{\perp} &=& \mu {g_L \over\mu_L},\qquad\label{gperp}\\
\rho_1 &=& \frac{\mu_L}{B},\qquad\label{rho1}\\
\rho_2 &=& \frac{\tilde{C}}{\sqrt{B \mu_L}}.\qquad\label{rho2}
\end{eqnarray}
Flow equations for $\rho_1(l)$ and $\rho_2(l)$ are given by:
\begin{eqnarray}
\frac{d \, \rho_1} {d \, l} &=& -\frac{3}{2} g_L \,\rho_1 \,(1 - \rho_2^2),
\label{rho1flow}\\
\frac{d \, \rho_2} {d \, l} &=& -\frac{3}{4} g_L\, \rho_2\, (1 -
\rho_2^2),\label{rho2flow}
\end{eqnarray} 
where mechanical stability requires $\rho_2 = \tilde{C}/\sqrt{B \,
\mu_L}<1$.  Below $5$ dimension, we expect elastic nonlinearities to
be relevant in the presence of quenched random strains and therefore
$g_L(l)$ to flow to a positive finite value. This together with
Eqs.\ref{rho1flow},\ref{rho2flow} implies that at long scales, ratios
$\rho_1(l)$ and $\rho_2(l)$ flow to zero, leading to a considerable
simplification of the flow equations for $g_L(l)\  (\neq 0)$ and
$g_{\perp}(l)$
\begin{eqnarray}
\frac{d \,g_L}{d \,l} &=& \epsilon {g_L} 
 - \frac{5{g_L}\left( 64{{g_L}}^2 + 176{g_L}{g_{\perp}} + 
      51{{g_{\perp}}}^2 \right) }{32
    \left( 8{g_L} + 3{g_{\perp}} \right) },\label{flow_gL}\\
 \frac{d \, g_{\perp}}{d \, l} &=& \epsilon g_{\perp} 
- \frac{ {g_{\perp}}\left( -64{{g_L}}^2 + 752{g_L}{g_{\perp}} + 
        261{{g_{\perp}}}^2 \right) }{32
    \left( 8{g_L} + 3{g_{\perp}} \right) }.\qquad\label{flow_gperp}
\end{eqnarray} 
For $d=3 < 5$, $g_L(l)$ and $g_{\perp}(l)$ indeed grow at long scales,
invalidating harmonic elasticity. Eqs.\ref{flow_gL},\ref{flow_gperp},
with flows displayed in Fig.\ref{flowdiagram}, have four fixed points,
Gaussian (G), Smectic (S), X, and Elastomer (E), that we list in Table
I.
\begin{table}[!hbt]
\label{fixedpoints_disorder}
\begin{tabular}{|c|c|c|c|c|c|c|c|c|}
\hline\hline Fixed point & $g_L*$ & $g_{\perp}*$
& $\eta_B$ & $\eta_L = \eta_{\tilde{C}}$ 
&$\eta_{\perp}$&$\eta_K$&$\eta_{\Delta}$\\
\hline G & 0 & 0 & 0 & 0&0&0&0 \\
S&$ \frac{4 \epsilon}{5}$&0& $0$ &
 $\frac{6 \epsilon}{5}$&0& $\frac{\epsilon}{5}$&$\frac{\epsilon}{10}$\\
X & 0 & $\frac{32 \epsilon}{87}$ &0&0&$\frac{2 \epsilon}{87}$
&$\frac{35 \epsilon}{87}$&$\frac{\epsilon}{58}$\\
E &$\frac{4 \epsilon}{263}$& $\frac{96 \epsilon}{263}$
 & 0 & $\frac{6 \epsilon}{263}$&$\frac{6 \epsilon}{263}$
&$\frac{106 \epsilon}{263}$&$\frac{5 \epsilon}{263}$\\
\hline
\end{tabular}
\caption{Fixed point couplings and $\eta$ exponents for heterogeneous nematic
elastomer. $\rho_{1,2}^*=0$ for all fixed points.}
\end{table}
\vspace{-0.2cm} It is clear from Eq.\ref{Hamiltonian_disorder} that
for a vanishing shear modulus $\mu=0$, transverse in-plane phonon
modes decouple from $u_z$. The remaining longitudinal mode is encoded
through in-plane density fluctuations $w_{ii}$, that can be integrated
out, leading to a finite shift in $B_z$, with the model reducing to
that of a randomly strained smectic\cite{RT_smectic}. It is reassuring
that this fixed point S, is precisely the same as that found in the
study of smectics in aerogel\cite{RT_smectic}. 
We are not aware of any physical system that is described by the fixed
point X, characterized by $g_L=0$.  However, generically elastomers
are characterized by a finite in-plane shear modulus $\mu$ and a
finite modulus $\mu_L$ ($g_L\neq0$), and, as can be seen from
Fig.\ref{flowdiagram}, S and X are unstable to these couplings. At
scales longer than nonlinear crossover lengths $\xi^{z, \perp}_{NL}$,
the system flows into a globally stable zero-temperature fixed point
E, controlled by elastomer heterogeneities and elastic nonlinearities.

Consequently, at long scales, nematic elastomer can be described by an
effectively harmonic but {\em nonlocal} elasticity, with all elastic
moduli, except the overall bulk modulus $B$ length-scale (wavevector)
dependent. Standard matching calculation\cite{RT_smectic} shows that
indeed
\begin{eqnarray}
K(\kv) &=& K_0 (k_{\perp} \xi^{\perp}_{NL})^{-\eta_K} 
f_K[(k_{\perp} \xi^{\perp}_{NL})^{\zeta}/k_{z} \xi^{z}_{NL}],
\label{K}\\
\mu(\kv) &=& \mu_0 (k_{\perp} \xi^{\perp}_{NL})^{\eta_{\perp}} 
f_{\mu}[(k_{\perp} \xi^{\perp}_{NL})^{\zeta}/k_{z} \xi^{z}_{NL}],
\label{mu}\\
\mu_L(\kv) &=& \mu_0 g^*_L/g^*_T (k_{\perp} \xi^{\perp}_{NL})^{\eta_{L}} 
f_{\mu_L}[(k_{\perp} \xi^{\perp}_{NL})^{\zeta}/k_{z}
\xi^{z}_{NL}],\qquad
\label{muL}\\
\tilde{C}(\kv) &=& \tilde{C}_0 (k_{\perp} \xi^{\perp}_{NL})^{\eta_{\tilde{C}}} 
f_{\tilde{C}}[(k_{\perp} \xi^{\perp}_{NL})^{\zeta}/k_{z}\xi^{z}_{NL}],
\label{C}\\
\Delta(\kv) &=& \Delta_0 (k_{\perp} \xi^{\perp}_{NL})^{-\eta_{\Delta}} 
f_{\Delta}[(k_{\perp} \xi^{\perp}_{NL})^{\zeta}/k_{z} \xi^{z}_{NL}],
\label{Delta}\\
B(\kv) &\approx& B_0,\label{B} 
\end{eqnarray} 
where the anisotropy exponent $\zeta = 2-(\eta_{\perp}+\eta_{K})/2$.
Above scaling functions $f_\alpha[v]$ ($\alpha=K, \mu, \mu_L, \Delta$)
are $v$-independent for large $v$ and scale as
$v^{\pm\eta_\alpha/\zeta}$ for a vanishing $v$, such that for
$k_\perp\rightarrow0$ limit, the elastic moduli are
$k_\perp$-independent. Since this predicts that at long length scale,
all elastic moduli are vanishingly small compared to the
wavevector-independent bulk modulus $B$, nematic elastomer is strictly
macroscopically {\em incompressible}, as advertised in the
introduction.

Furthermore, from definitions of the coupling $g_L$ and $g_{\perp}$ we
find that the ratio between shear moduli $\mu$ and $\mu_L$ approaches
a universal value $g^*_{\perp}/g^*_L = 24$ at the fixed point E. This,
together with a straightforward analysis and Eq.\ref{gperp} predicts
that a uniform strain $u_{xx}$\cite{comment_n0} leads to
\begin{eqnarray}
u_{yy} &=& \frac{\mu^R-4 \mu^R_L}{\mu^R+4\mu^R_L}\; u_{xx} 
\rightarrow\frac{5}{7} u_{xx},\label{uyy}\\ 
u_{zz} &=& - \frac{2 \mu^R}{\mu^R+4\mu^R_L}\; u_{xx} 
\rightarrow -\frac{12}{7} u_{xx},\label{uzz}
\end{eqnarray} 
with negative and positive universal Poisson ratios.

Another fascinating implication of Eqs.\ref{K}-\ref{B} is a strictly
nonlinear stress-strain response down to an arbitrarily weak stress
$\sigma_{zz} < \sigma_{NL}\equiv
K/\xi_{NL,\perp}^2$\cite{semisoft}. To show this we note that
$\sigma_{zz}$ cuts off the singular $k$ dependence
(Eqs.\ref{K}-\ref{Delta}) of elastic moduli on scales longer than
$\xi_\sigma^\perp=(K\xi_{NL,\perp}^{-\eta_K})^\nu \sigma_{zz}^{-\nu}$,
$\nu=1/(2-\eta_K)$, thereby replacing it by a singular $\sigma_{zz}$
dependence
$\sim(\sigma_{zz}/\sigma_{NL})^{\eta_\alpha\nu}$.\cite{RT_other} In
particular we find a non-Hookean response $u_{zz}\sim
\sigma_{zz}^{1/\delta}$, $1/\delta = 1-\eta_{L}/(2-\eta_K)$, that
within (inaccurate) $\epsilon=2$-expansion gives $\delta=157/151$
unimpressively close to $1$. 

Our results rely on the assumption of a stable long-range nematic
order\cite{stableN}, requiring convergent orientational fluctuations
\begin{equation} \overline{\langle |\delta \nv|^2 \rangle} \approx
\overline{\langle |\partial_{\perp} u_z|^2 \rangle} \sim
L^{\eta_K+\eta_{\perp}-2}.  \end{equation} Thus a necessary condition for
stability of nematic order is $\eta_K+\eta_{\perp}-2<0$, equivalent to the
requirement on the anisotropy exponent $\zeta >1$. This is satisfied for $d >
d_{lc} = 17/56$, suggesting a stability of nematic order for 3d elastomers.

Experimentally, elastomers crosslinked in the isotropic exhibit a
polydomain nematic order with a typical micron size correlation
length\cite{heterogeneity,stableN}. This can be reconciled with the
above prediction of the $\epsilon$-expansion by heterogeneity that is
not weak, or by appealing to glassy dynamics that prevents a full
equilibration on experimentally relevant time scales. This hypothesis
can tested by cooling in the presence of a strong aligning field.

To sum up, we have studied nematic elastomers, taking into account
their heterogeneity, and have shown that it leads to many striking
elastic properties, that should be readily experimentally testable.

We thank T. Lubensky and J. Toner for discussion and acknowledge
support by the NSF MRSEC DMR98-09555 (LR, XX), the A. P. Sloan and the
David and Lucile Packard Foundations (LR), and the University of
Colorado Faculty Fellowship (LR). We also thank Harvard Department of
Physics, where part of this work was done, for hospitality.


\vspace{-0.5cm}
 \begin{thebibliography}{99}    
\vspace{-0.5cm}   
 \bibitem{review_elastomers}M.~Warner and E.~M.~Terentjev,{\em Prog.
     Polym. Sci.}  {\bf 21}, 853(1996); E. M. Terenjev, {\em J. Phys.
     Cond. Mat.}  {\bf 11}, R239(1999).
   
 \bibitem{semisoft} We focus on ``ideal'' gels that exhibit a
   statistically isotropic and homogeneous high-temperature phase, in
   contrast to the so-called ``semisoft'' gels that are intrinsically
   anisotropic.
   
 \bibitem{GL} L.~Golubovic and T.~C.~Lubensky, {\em \prl} {\bf 63},
   1082 (1989).
   
 \bibitem{LMRX} T.~C.~Lubensky, R.~Mukhopadhyay, L.~Radzihovsky, and
   X.~Xing, {\em condmat-0112095}.
   
 \bibitem{Olmsted}Peter D. Olmsted. {\em J. Phys. II(France)}, {\bf
     4}, 2215 (1994).
   
 \bibitem{softmodes} H. Finkelmann, {\em et. al},
%I. Kundler, E.M. Terentjev, and M. Warner, 
{\em J. Phys. II} {\bf 7}, 1059 (1997); 
G.C. Verwey, {\em et. al},
%M. Warner, and E.M. Terentjev, 
J. {\em Phys. II (France)} {\bf 6}, 1273-1290 (1996).
%M. Warner, {\em J. Mech. Phys. Solids} {\bf 47}, 1355 (1999).

% \bibitem{deGennes}P.G.~de Gennes in {\em Liquid Crystals
%                 of One and Two-Dimensional Order}, p. 231, edited by
%               W. Helfrich and G.Heppke (Springer, New York, 1980)
   
 \bibitem{XRunpublished} X.~Xing and L.~Radzihovsky, unpublished.
   
 \bibitem{elastomer_thermal} X.~Xing and L.~Radzihovsky; T.~C.~Lubensky,
   and O.~Stenull, unpublished.
   
 \bibitem{heterogeneity} See e.g., N.~Uchida, {\em \pre} {\bf 62},
 5119 (2000).
   
 \bibitem{comment_n0} For $u_{xx}>0$, this must be done in a clamped
   geometry so as to ensure that the overall uniaxial nematic axis
   ${\bf\hat{n}}_0$ does not reorient. For $u_{xx}<0$ no such
   percussions are necessary.

 \bibitem{Gingras} M.~J.~P.~Gingras, unpublished.
   
 \bibitem{RT_smectic} L.~Radzihovsky and J.~Toner, {\em \prl} {\bf 78}, 4414
   (1997); {\em \prb} {\bf 60} 206 (1999); T.~Bellini, L.~Radzihovsky,
   J.~Toner, N.~A.~Clark, {\em Science} {\bf 294} 1074 (2001).
   
 \bibitem{Feldman} D. E. Feldman, {\em \prl} {\bf 84}, 4886 (2000).
%\prb {\bf 61}, 382 (2000).
   
 \bibitem{membranes} L. Radzihovsky and D. R. Nelson, {\em \pra} {\bf 44},
   3525 (1991);  D. Morse and T. C. Lubensky, ibid. {\bf 46}, 1751 (1992).
   
 \bibitem{RT_other} B. Jacobsen, {\em et. al}, {\em \prl} {\bf 83}, 1363
   (1999); K.  Saunders, {\em et. al}, ibid. {\bf 85}, 4309 (2000); L.
   Radzihovsky, {\em et. al}, ibid. {\bf 87}, 027001 (2001).
   
 \bibitem{stableN} If in fact the nematic order is unstable
   (notwithstanding $\epsilon$-expansion prediction), or experiments
   are done on not fully equilibrated multi-domain samples, we expect
   our results to continue to hold out to an orientational length
   $\xi_O$ ($\gg\xi_{NL}$ for $\Delta\rightarrow 0$
   \cite{RT_smectic}), beyond which we expect elastomers to
   exhibit conventional elasticity.

 \bibitem{no_rf} Because elastomer heterogeneity does not break
   translational symmetry of the target space, unlike liquid crystals
   confined to aerogel studied in
   Refs.\onlinecite{RT_smectic,RT_other}, here, no random phonon-field
   pinning appears.

\end{thebibliography}
\end{document}